\documentclass[9pt,twocolumn,twoside]{osajnl}

\newcommand{\betabold}{\mbox{\boldmath$\beta$}}

\newcommand{\alphabold}{\mbox{\boldmath$\alpha$}}

\newcommand{\one}{\mbox{\boldmath$1$}}
\DeclareMathOperator*{\argmin}{arg\,min}
\DeclareMathOperator*{\prox}{\mathrm{prox}}

\journal{ao} 

\setboolean{shortarticle}{false} 

\title{Random sub-Nyquist polarimetric modulator}

\author[1,2]{A. Asensio Ramos}

\affil[1]{Instituto de Astrof\'{\i}sica de Canarias, 38205, La Laguna, Tenerife, Spain}
\affil[2]{Departamento de Astrof\'{\i}sica, Universidad de La Laguna, E-38205 La Laguna, Tenerife, Spain}

\affil[*]{Corresponding author: aasensio@iac.es}

\dates{Compiled \today}

\ociscodes{(120.2130) Ellipsometry and polarimetry; (120.5410) Polarimetry}

\doi{\url{http://dx.doi.org/10.1364/ao.XX.XXXXXX}}

\begin{abstract}
We show that it is possible to measure polarization with a polarimeter that gets rid of the seeing while 
still measuring at a frequency well below that of the seeing.
We study a standard polarimeter made of two retarders and a beamsplitter. The retarders are modulated at $\sim 500$ Hz, a frequency
comparable to that of the variations of the refraction index in the Earth atmosphere, what is
usually termed as seeing in astronomical observations. However, we assume 
that the camera is slow, so that our measurements are time integrations of these modulated signals. In order to recover
the time variation of the seeing and obtain the Stokes parameters, we use the theory of compressed sensing to solve the demodulation
by impose a sparsity constraint on the Fourier coefficients of the seeing.
We demonstrate the feasibility of this sub-Nyquist polarimeter using numerical simulations, both in the case without 
noise and with noise. We show that a sensible modulation scheme is obtained by randomly changing the fast axis of the
modulators or their retardances in specific ways. We finally demonstrate that the value of the Stokes parameters can be recovered with
great precision at almost maximum efficiency, although it slightly degrades when the signal-to-noise ratio of the observations increase, a consequence
of the multiplexing under the presence of photon noise.
\end{abstract}

\setboolean{displaycopyright}{true}

\begin{document}

\maketitle
\thispagestyle{fancy}
\ifthenelse{\boolean{shortarticle}}{\abscontent}{}

\section{Introduction}
The lack of detectors sensitive to the polarization state of the light has forced us
to use linear measurement schemes (also known as modulation or multiplexing) for Stokes polarimetry. The incoming Stokes
parameters are encoded into the intensity of the light and several such linear measurements are 
carried out. The ensuing Stokes parameters are then recovered by solving the corresponding determined or
overdetermined linear system of equations using the inverse or the Moore-Penrose pseudo-inverse \citep{deltoro_collados00}.
One of the key assumptions that this scheme needs in order to work properly is that the target Stokes parameters
do not change during the time it takes to carry out one of the modulation cycles (a cycle is defined as the
time necessary to make the measurements so that we end up with a solvable linear system).

This assumption is clearly broken under the presence of seeing\footnote{For the sake of precision,
astronomical seeing refers to perturbation produced in astronomical images because of the 
spatial and time variation of the refractive index of the air on the Earth's atmosphere.} because it contains frequencies 
well above those that can be reached with our current cameras. As a consequence, the Stokes parameters
entering into our polarimeter do change inside one of the measurement cycles. This translates into
the presence of the well-studied seeing-induced cross-talk \citep{lites87,judge04,casini12}, which leads to
spurious signals after the demodulation.

Several possibilities have been devised in the past to reduce this seeing-induced cross-talk. The most
obvious one is to use very fast cameras so that the modulation scheme can be made faster than the variations
of the seeing. An example of this is the very successful instrument ZIMPOL \citep[Zurich IMaging POLarimeter;][]{povel95}.
This polarimeter has a custom-made camera that carries out the modulation electronically by charge displacement
and can work at kHz frequencies. This high speed freezes the seeing and no seeing-induced crosstalk appears during
the modulation-demodulation process. 

\begin{figure*}
\includegraphics[width=\textwidth]{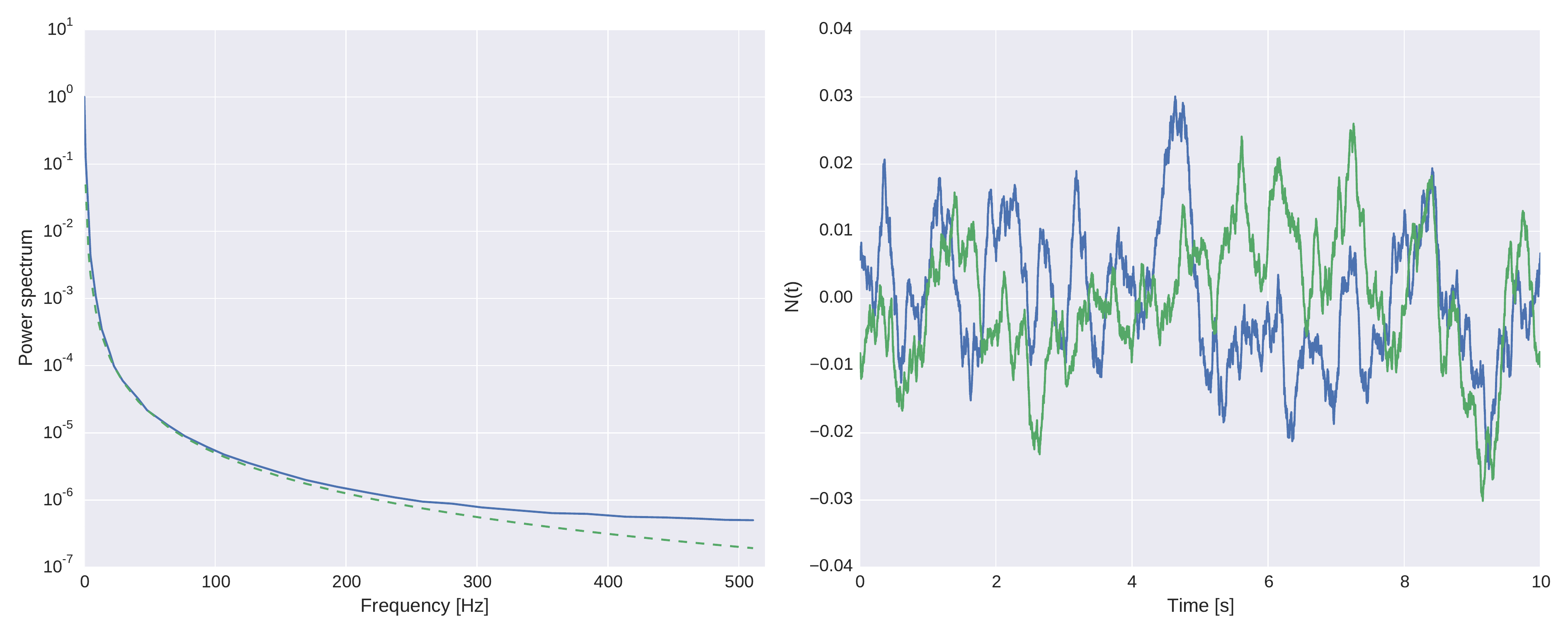}
\caption{Left panel: measured power spectrum of the seeing \cite{lites87} (blue), together with the curve $f^{-2}$ (dashed green). Right panel: two
realizations of such Gaussian random process, both with unit variance and zero mean. The time interval
$\Delta t=1$ ms.}
\label{fig:powerSpectrum}
\end{figure*}

Another possibility to reduce the effect of seeing is to use double beams, 
taking advantage of beamsplitters, that provide two orthogonally polarized beams. Two cameras (or two portions of the
same camera) are used to detect the two orthogonal beams, which are observed strictly simultaneously. This  
cancels out the effect of seeing at first order but introduces other problems related to the potentially different optical paths
of the two beams or to the different flat-fielding effects of the two beams. It is then possible to combine the
time modulation with the presence of a double-beam to again cancel out at first order these flat-fielding 
problems. The success of double-beam polarimetry coupled with slow modulation is so widespread that this idea is now 
at the heart of almost every single operative polarimeter.
The ideal situation would be then to have a double-beam polarimeter that can modulate at kHz frequencies, with a camera
that can also measure at these speeds. This will allow us to properly demodulate a full cycle before the Stokes parameters
change due to the presence of seeing. This is currently a technological challenge. 
Current piezoelastic and electro-optic modulators are now able to carry out the 
polarimetric modulation 
at kHz rates without difficulties. Ferroelectric Liquid Crystal (FLC) are still lagging behind
in terms of modulation frequency but very fast Liquid Crystal Variable Retarders (LCVR) are now
commercially available. On the other hand, commercial cameras with enough sensitivity cannot go so fast.

We propose in this paper a conceptual idea for a polarimeter in which modulation happens at
kHz frequencies but the measurements are done at much reduced speeds. The immediate consequence is that the linear system
produced in the modulation is underdetermined, so that an infinite number of solutions exist. A direct application of the theory 
of band-limited signals demonstrates that it is impossible to reconstruct back the signal without introducing any
additional constraint. However, it is also true that the recent theory of compressed sensing \citep[CS;][]{candes06,donoho06} shows that
the Nyquist limit is too restrictive if one is able to impose a sparsity prior on the signal. Inspired by recent results \cite{tropp10}, we
demonstrate that such a sub-Nyquist\footnote{We refer to this polarimeter as sub-Nyquist because 
our measurements are below the Nyquist limit of the seeing.} polarimeter is feasible. We hope that
the idea presented here can become reality in the near future thanks to advances in
optical instrumentation.

\section{Theory of a random sub-Nyquist polarimetric modulator}
We present in this section the theoretical and numerical tools that are necessary to deal with a sub-Nyquist polarimeter. We describe
the seeing as a time-correlated random Gaussian process and we show the compressibility of the Fourier
transform of the seeing process. This allows us to transform the problem into an instance of the theory of 
CS, which allows us to regularize the solution by imposing a sparsity constraint. 

\subsection{Seeing random process}
In this section, we follow the approach of \cite{lites87}, \cite{judge04} and \cite{casini12}
to describe the effect of seeing on the measured Stokes parameters.
In a seeing-free situation, the measured Stokes parameter $S_i(x,y,t)$, with 
$i=1,\ldots,4$ (for Stokes $I$, $Q$, $U$ and $V$) at a position $(x,y)$ and for time $t$ would be equal to the 
unperturbed Stokes parameter :
\begin{equation}
S_i(x,y,t) = R_i(x,y,t),
\end{equation}
where $R_i(x,y,t)$ is the Stokes vector unaffected by seeing that arrives to the upper
layers of the Earth atmosphere. Under the presence of seeing, the observed Stokes parameter 
does not correspond exactly to the same $(x,y)$ position on the Sun, so that the equation describing the observation is really
\begin{equation}
S_i(x,y,t) = R_i(x',y',t).
\end{equation}
Under the approximation that the seeing is not very large, we can do a Taylor expansion 
to first order and we find that
\begin{equation}
S_i(x,y,t) = R_i(x,y,t) + \nabla R_i(x,y,t) \cdot \mathbf{s}(t),
\end{equation}
where $\mathbf{s}(t)$ is the displacement at each time produced by the curvature in the wavefront and $\nabla R_i(x,y,t)$
is the spatial gradient of the Stokes profiles emerging from the solar surface. Under the
assumption that $\mathbf{s}(t)$ has no preferred direction with time (seeing is isotropic), the previous equation can
be expresssed as \citep{lites87,judge04,casini12}:
\begin{equation}
S_i(t) = R_i \left[ 1 + \beta_i N(t)\right],
\label{eq:seeingEffect}
\end{equation}
where $N(t)$ is a Gaussian process with zero mean and unit variance with a power 
spectrum $P_N(\nu)$ defined by the seeing power spectrum \cite[see Fig. 1 of][]{lites87}.
If the random process is assumed to be normalized to unit area (we remind that the area of the power spectrum
is equal to the variance of the random process), so that
\begin{equation}
\int_0^\infty P_N(\nu) d\nu = 1,
\end{equation}
then $\beta_i$ corresponds to the standard deviation of the seeing process. The left panel of Fig.
\ref{fig:powerSpectrum} shows the power spectrum extracted from \cite{lites87}, while the right
panel displays two different realizations of a Gaussian process with such power spectrum. The time
interval $\Delta t$ of the discretized process is 1 ms, short enough to accommodate all frequencies.

Making Eq. (\ref{eq:seeingEffect}) explicit for all Stokes parameters, we have that
\begin{align}
I(t) &= I_0 \left[ 1+ \beta_I N(t) \right] \\
Q(t) &= Q_0 \left[ 1+ \beta_Q N(t) \right] \\
U(t) &= U_0 \left[ 1+ \beta_U N(t) \right] \\
V(t) &= V_0 \left[ 1+ \beta_V N(t) \right],
\end{align}
where $\mathcal{I}=(I_0,Q_0,U_0,V_0)$ are the solar values for the Stokes parameters, that we 
assume fixed during the observation. This assumption might be relaxed in the future if we assume 
that there is a slow variation with time, but more studies on this direction are necessary.
Additionally, and for simplicity of notation, we also define the vector $\betabold=(\beta_I,\beta_Q,\beta_U,\beta_V)$.

\subsection{Modulation}
The difficulty in dealing with the seeing in normal polarimeters 
is that frequencies up to 500 Hz are present, as shown in Fig. \ref{fig:powerSpectrum}. Therefore, if one wants 
to freeze the seeing, the whole modulation-demodulation process has to
take place roughly at kHz rates. As noted in the introduction, except for a few exceptions, current
cameras are not able to measure at these rates so that all measurements
are affected by seeing due to the finite integration times.

In this section, we describe a polarimeter that modulates at kHz rates (therefore
freezing the seeing) but measurements are done much slower. Given that our cameras are slow (even 
an order of magnitude slower) and we are not able to observe the modulated 
signals at such high speeds, each measurement that we do in 
the camera is the result of summing up the signal that is modulated by the 
seeing and the polarimetric modulator. 

In a double beam polarimeter, the instantaneous output of the polarimeter for the two beams is given by
\begin{align}
S_1(t) &= M_1(t) I(t) + M_2(t) Q(t)+ M_3(t) U(t) + M_4(t) V(t) \nonumber \\
S_2(t) &= M_1(t) I(t) - M_2(t) Q(t)- M_3(t) U(t) - M_4(t) V(t),
\end{align}
where $I(t)$, $Q(t)$, $U(t)$ and $V(t)$ are the instantaneous values of the Stokes parameters arriving to
the polarimeter, while $M_1(t)$, $M_2(t)$, $M_3(t)$ and $M_4(t)$ are the known modulation introduced by the polarimeter, that
are changed with a period $\Delta t$. For analyzing signals perturbed with seeing, we need $\Delta t$ to be in the
millisecond range.

Since the camera is much slower than the seeing, our measurements at step $j$ are 
time integrals in the time interval $[t_\mathrm{start}^j,t_\mathrm{end}^j]$ of length $\Delta T=t_\mathrm{start}^j-t_\mathrm{end}^j$:
\begin{align}
S_1^j &= \int_{t_\mathrm{start}}^{t_\mathrm{end}} \mathrm{d}t \left[ M_1(t) I(t) + M_2(t) Q(t) + M_3(t) U(t) + M_4(t) V(t) \right] \nonumber \\
S_2^j &= \int_{t_\mathrm{start}}^{t_\mathrm{end}} \mathrm{d}t \left[ M_1(t) I(t) - M_2(t) Q(t) - M_3(t) U(t) - M_4(t) V(t) \right],
\end{align}
where $t_\mathrm{start}^j$ is the time at which the camera starts integrating and $t_\mathrm{end}^j$ 
is the time at which the integration is finalized. Note that $M=\Delta T / \Delta t \gg 1$ is the
number of time steps in each camera exposition. Typically, for our current technologies, $M \sim 10$ or larger, corresponding
to integrations in the range of tens of milliseconds.

The previous equations can be discretized at the $N$ times $t_i$, so that $t_{i+1}-t_i=\Delta t$ under the assumption that the 
modulation and the Stokes parameters are kept fixed during these intervals. The process of measurement at step $j=0 \ldots N_\mathrm{meas}-1$ 
is given by:
\begin{align}
S_1^j &= \sum_{i=jM}^{j(M+1)-1}  \left[ M_1(t_i) I(t_i) + M_2(t_i) Q(t_i) \right. \nonumber \\
&+ \left.  M_3(t_i) U(t_i) + M_4(t_i) V(t_i)\right] \\
S_2^j &= \sum_{i=jM}^{j(M+1)-1} \left[ M_1(t_i) I(t_i) - M_2(t_i) Q(t_i) \right. \nonumber \\
&- \left. M_3(t_i) U(t_i) - M_4(t_i) V(t_i)\right].
\end{align}

The obvious advantage of the discretized equations is that they can be easily written in matrix form. To this end,
we build the matrices $\mathbf{M}_1$, $\mathbf{M}_2$, $\mathbf{M}_4$ and $\mathbf{M}_4$ of size $N_\mathrm{meas} \times N$ that represent
the modulation states at all time steps. Each row of these matrices will contain the value of $M_k(t_i)$ for each measurement $j$ arranged
on the following manner:
\begin{equation}
\mathbf{M}_k = \left[
\begin{array}{ccccccc}
M_k(t_1) & \cdots & M_k(t_M) & 0 & 0 & 0 & \cdots \\
0 & 0 & 0 & M_k(t_{M+1}) & \cdots & M_k(t_{2M}) & \cdots \\
0 & 0 & 0 & 0 & 0 & 0  \\
\cdots & \cdots & \cdots & \cdots & \cdots & \cdots & \cdots 
\end{array}
\right].
\end{equation}
This is a sparse fat matrix in which each row $j$ is zero except for the $M$ values $M_k(t_{jM}),\ldots,M_k(t_{j(M+1)-1})$. Using 
the matrix notation, the measurement process is then given by:
\begin{align}
\mathbf{S}_1 &= \mathbf{M}_1 \mathbf{I} + \mathbf{M}_2 \mathbf{Q} + \mathbf{M}_3 
\mathbf{U} + \mathbf{M}_4 \mathbf{V} \nonumber \\
\mathbf{S}_2 &= \mathbf{M}_1 \mathbf{I} - \mathbf{M}_2 \mathbf{Q} - \mathbf{M}_3 
\mathbf{U} - \mathbf{M}_4 \mathbf{V},
\label{eq:linearSystemMeasurements}
\end{align}
where $\mathbf{S}_i$ are column vectors of length $N_\mathrm{meas}$, while $\mathbf{I}$, $\mathbf{Q}$, $\mathbf{U}$ and $\mathbf{V}$ are column vectors of length $N$.

The previous equations consider the fast polarimeter that integrates at the rate of the seeing, if one assumes that the size of 
$\mathbf{S}_i$ equals that of $\mathbf{I}$, $\mathbf{Q}$, $\mathbf{U}$ and $\mathbf{V}$ (in other words, $N_\mathrm{meas}=N$). In this case, 
Eq. (\ref{eq:linearSystemMeasurements}) represents a solvable linear system of equations. This linear system has more 
equations than unknowns (because of the double-beam strategy) and can be solved efficiently in the least-square sense by using 
the Penrose pseudo-matrix \citep{deltoro_collados00}.

\subsection{Sub-Nyquist demodulation}
In the more difficult case of a slow camera, the size of $\mathbf{S}_i$ is much smaller than that of
the Stokes parameters $\mathbf{I}$, $\mathbf{Q}$, $\mathbf{U}$ and $\mathbf{V}$. 
Consequently, the linear system of Eq. (\ref{eq:linearSystemMeasurements}) is underdetermined 
and a unique solution does not exist.

However, it is still possible to solve the problem if we impose a prior on the signal. In
our case, we will exploit the fact that the seeing has a power spectrum that falls roughly as 
$1/f^2$ (see the dashed green line of Fig. \ref{fig:powerSpectrum}). This means that the amplitude of the coefficients $\alpha_i$ of the 
Fourier expansion of the seeing Gaussian random process $N(t)$ fall roughly as $\sim 1/f$. As a consequence, the seeing random process can be considered
to be compressible or weakly sparse in the Fourier basis\footnote{A signal is compressible or weakly sparse in a basis if the coefficients
of the expansion in that basis fulfill $|\alpha_i| \leq C i^{-1/r}$, with $C$ a constant and $r \sim 1$.}. This makes it possible to take advantage of the CS theory for 
compressible signals to estimate the high-frequency variation of the seeing using low frequency camera expositions.

To this end, we write the discretized Stokes parameters in terms of the Fourier coefficients as:
\begin{align}
\mathbf{I} &= I_0 \left( \one + \beta_I \mathbf{F}^{-1} \alphabold \right) \nonumber \\
\mathbf{Q} &= Q_0 \left( \one + \beta_Q \mathbf{F}^{-1} \alphabold \right) \nonumber \\
\mathbf{U} &= U_0 \left( \one + \beta_U \mathbf{F}^{-1} \alphabold \right) \nonumber \\
\mathbf{V} &= V_0 \left( \one + \beta_V \mathbf{F}^{-1} \alphabold \right),
\label{eq:discretizedStokes}
\end{align}
where $\alphabold=(\alpha_1,\alpha_2,\ldots,\alpha_N)^T$ is a column vector containing the Fourier coefficients of the compressible seeing random process, 
$\one=(1,1,\ldots,1)^T$ is the unit column vector of length $N$, while $\mathbf{F}^{-1}$ is the inverse discretized Fourier matrix, that
can be applied efficiently using the inverse fast Fourier transform (FFT).
We note that the system of Eqs. (\ref{eq:linearSystemMeasurements}) becomes a nonlinear set of equations for the unknowns
$(\mathcal{I},\betabold,\alphabold)$ when using Eqs. (\ref{eq:discretizedStokes}). This nonlinearity
appears because we are introducing some prior information about the time variation of the Stokes parameters and it is 
precisely this prior information the one that allows us to solve the inverse problem.

A solution to the system of Eqs. (\ref{eq:linearSystemMeasurements}) can then be obtained in the least-squares
sense by optimizing the $\ell_2$ norm\footnote{The $\ell_q$-norm
is given by: $\Vert \mathbf{x} \Vert_p = \sum_i |x_i|^p$, with $p\geq0$.} of the residuals. The compressibility constraint can be imposed by forcing the $\ell_1$-norm
of the Fourier coefficients $\alphabold$ to be small \citep{candes06}. With this in mind, a particular solution 
\citep[which is the correct solution with a large probability;][]{candes06}
to the system of Eqs. (\ref{eq:linearSystemMeasurements}) can be obtained by solving the following problem:
\begin{equation}
\argmin_{\alphabold,\betabold,\mathcal{I}} \Vert \mathbf{S}_1 - \mathbf{S}_1^\mathrm{obs} \Vert_2^2 + \Vert \mathbf{S}_2 - \mathbf{S}_2^\mathrm{obs} \Vert_2^2 + \lambda \Vert \alphabold \Vert_1,
\label{eq:optimization}
\end{equation}
where $\lambda$ is a regularization parameter, $\mathbf{S}_1^\mathrm{obs}$ and $\mathbf{S}_2^\mathrm{obs}$ are the observed modulated signals, and
\begin{align}
\mathbf{S}_1 &= \left[ \mathbf{M}_1 I_0 \left(\mathbf{1}+\beta_I \mathbf{F}^{-1} 
\alpha \right) + \mathbf{M}_2 Q_0 \left(\mathbf{1}+\beta_Q \mathbf{F}^{-1} \alpha \right) \right. \nonumber \\
& \left. + \mathbf{M}_4 U_0 \left(\mathbf{1}+\beta_U \mathbf{F}^{-1} \alpha \right) + \mathbf{M}_4 V_0 \left(\mathbf{1}+\beta_V \mathbf{F}^{-1} \alpha \right) \right] \nonumber \\
\mathbf{S}_2 &= \left[ \mathbf{M}_1 I_0 \left(\mathbf{1}+\beta_I \mathbf{F}^{-1} 
\alpha \right) - \mathbf{M}_2 Q_0 \left(\mathbf{1}+\beta_Q \mathbf{F}^{-1} \alpha \right) \right. \nonumber \\
& \left. - \mathbf{M}_4 U_0 \left(\mathbf{1}+\beta_U \mathbf{F}^{-1} \alpha \right) - \mathbf{M}_4 V_0 \left(\mathbf{1}+\beta_V \mathbf{F}^{-1} \alpha \right) \right].
\end{align}

The demodulation of the polarimetric measurements is carried out by optimizing Eq. (\ref{eq:optimization}) with respect 
to $\alphabold$, $\mathcal{I}$ and $\betabold$. This optimization is not an easy task because the merit
function is a non-convex function of the parameters (because of the sparsity constraint) 
and, additionally, these parameters enter nonlinearly into the merit function. To this end, we 
use an alternating optimization method that has been empirically very successful for solving complex problems. 
When $\mathcal{I}$ and $\betabold$ are kept fixed, the optimization of Eq. (\ref{eq:optimization}) becomes
linear in $\alphabold$. The same happens when we fix $\alphabold$ and $\mathcal{I}$ and optimize with respect to $\betabold$ and
also when $\alphabold$ and $\betabold$ are kept fixed and optimize with respect to $\mathcal{I}$. Therefore, our optimization is 
done by iteratively solving these simpler linear problems. In our experience, we find that carrying out the optimization
with respect to $\mathcal{I}$ and $\betabold$ only every 10 iterations gives systematically robust results.

\subsubsection{Optimization with respect to $\alphabold$}
Once $\mathcal{I}$ and $\betabold$ are kept fixed, the optimization of Eq. (\ref{eq:optimization}) is a standard compressed 
sensing problem for $\alphabold$. This can be efficiently solved using proximal algorithms \citep{parikh_boyd14} like the 
fast iterative shrinkage-thresholding algorithm \cite[FISTA;][]{beck_teboulle09}. This algorithm is suited for the
solution of problems where the merit function is given by 
\begin{equation}
f(\alphabold) = g(\alphabold) + h(\alphabold),
\end{equation}
where $g(\alphabold)$ is a convex function and $h(\alphabold)$ contains non-convex constraints. In our case, 
$g(\alphabold)=\Vert \mathbf{S}_1 - \mathbf{S}_1^\mathrm{obs} \Vert_2^2 + \Vert \mathbf{S}_2 - \mathbf{S}_2^\mathrm{obs} \Vert_2^2 $,
while $h(\alphabold)=\lambda \Vert \alphabold \Vert_1$ and the algorithm proceeds as follows:

\begin{algorithm}
\caption{FISTA algorithm}\label{alg:fista}
\begin{algorithmic}[1]
\Procedure{FISTA}{$L, g, \lambda$}
\State $\alphabold_0 = \mathbf{0}$, $\mathbf{y}_1= \alphabold_0$, $t_1=1$
\While{not converged}
\State $\alphabold_k = \prox_{\ell_1,\lambda} \left[\mathbf{y}_k - \frac{1}{L} \nabla g \right] $
\State $t_{k+1}= \frac{1}{2} \left(1+\sqrt{1+4t_k^2} \right)$
\State $\mathbf{y}_{k+1}=\alphabold_k + \left( \frac{t_k-1}{t_{k+1}} \right) \left( \alphabold_k - \alphabold_{k-1} \right)$.
\EndWhile\label{euclidendwhile}
\State \textbf{return} $\alphabold_k$
\EndProcedure
\end{algorithmic}
\end{algorithm}

The algorithm is started with a first estimation of the Fourier coefficients $\alphabold_0$, which we choose
to be all zeros. The symbol $L$ stands for the Lipschitz constant of $g(\alphabold)$, that can be computed from the spectral norm of the 
$\mathbf{M}_k$ matrices. We only need a lower limit for $L$ and the algorithm will always converge if $L$ is larger than the
correct Lipschitz constant.

This iteration is an accelerated version of the gradient descent followed by the application of the 
proximal operator for the $\ell_1$ norm. In our case of an $\ell_1$ sparsity constraint, the proximal operator is the smooth thresholding
operator \citep{parikh_boyd14}, which is given by:
\begin{equation}
\mathrm{prox}_{\ell_1,\lambda}(\mathbf{x}) = \mathrm{sign}(\mathbf{x}) (|\mathbf{x}|-\lambda)_+,
\end{equation}
where $(\cdot)_+$ denotes the positive part. The value of the threshold $\lambda$ has to be set in
advance. In our experience, a suitable value turns out to be the expected power of the noise (which in our case
turns out to be $\lambda \sim 10^{-6}$).

The calculation of the gradient of $g(\alphabold)$ with respect to $\alphabold$ can be done analytically taking into account that:
\begin{align}
\nabla_{\alphabold} \Vert \mathbf{S}_1 - \mathbf{S}_1^\mathrm{obs} \Vert_2^2 &= -2 \mathbf{F} \left[ I_0 \mathbf{M}_1^T + Q_0 \mathbf{M}_2^T + U_0 \mathbf{M}_3^T + V_0 \mathbf{M}_4^T
\right] \nonumber \\
& \times 
\left[\mathbf{S}_1 - \mathbf{S}_1^\mathrm{obs} \right] \nonumber \\
\nabla_{\alphabold} \Vert \mathbf{S}_2 - \mathbf{S}_2^\mathrm{obs} \Vert_2^2 &= -2 \mathbf{F} \left[ I_0 \mathbf{M}_1^T - Q_0 \mathbf{M}_2^T - U_0 \mathbf{M}_3^T - V_0 \mathbf{M}_4^T
\right] \nonumber \\
& \times 
\left[\mathbf{S}_2 - \mathbf{S}_2^\mathrm{obs} \right],
\end{align}
where $\mathbf{F}$ is the discretized Fourier matrix, that can be easily applied using the FFT.

\begin{figure*}
\includegraphics[width=\textwidth]{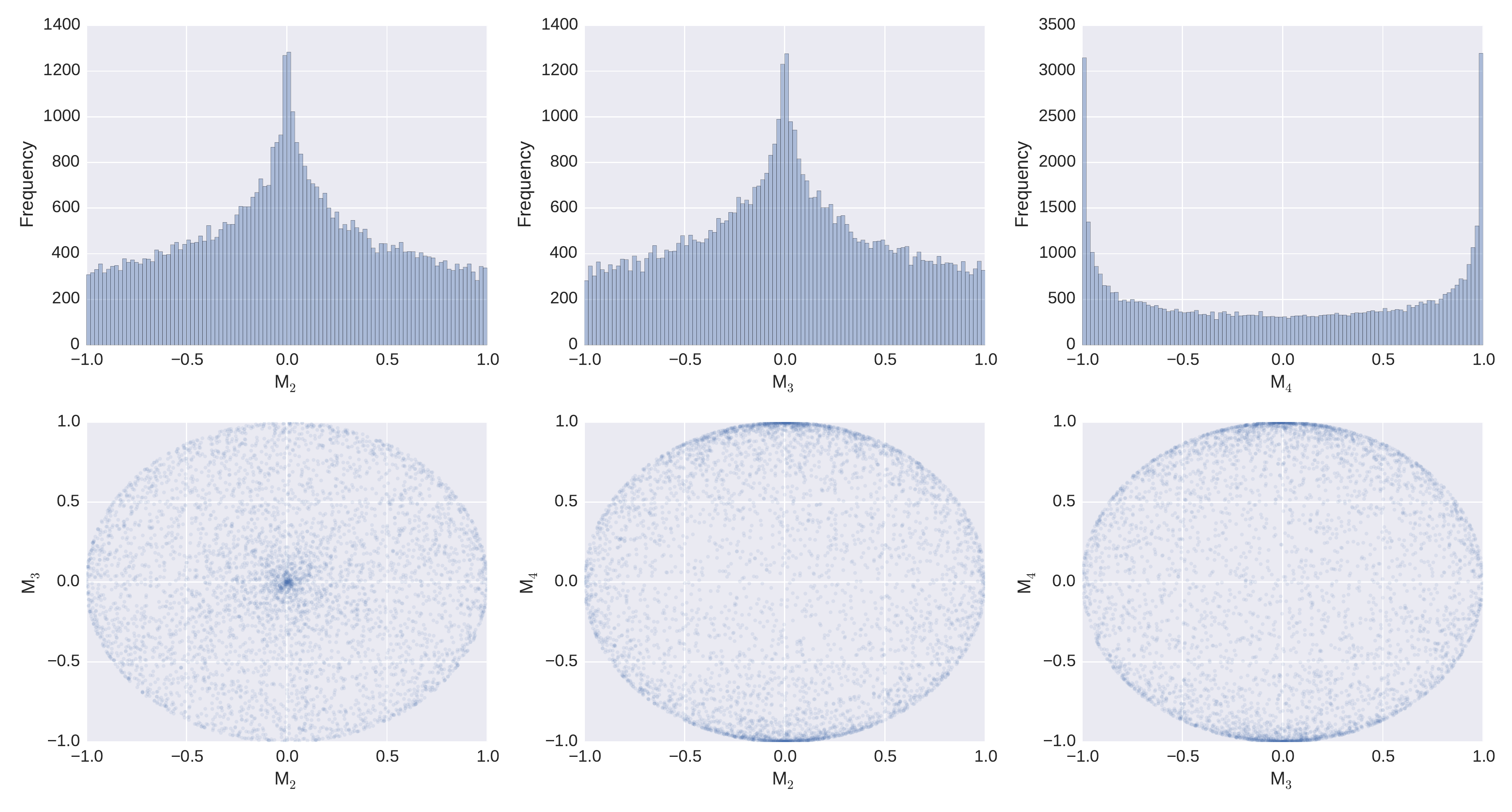}
\caption{Statistical properties of the elements of the modulation matrix obtained from Eq. (\ref{eq:modulationFLC})
when the fast-axis angles of the $\lambda/2$ and $\lambda/4$ are chosen randomly. This corresponds to a polarimeter
made of two FLCs. Note that $M_1(t)=1$ always. The upper panels display the marginal distributions, while the lower panels
display the joint distributions.}
\label{fig:modulation}
\end{figure*}

\subsubsection{Optimization with respect to $\mathcal{I}$}
When $\alphabold$ and $\betabold$ are kept fixed, Eq. (\ref{eq:optimization}) is a linear problem with respect to $\mathcal{I}$, whose
solution can be computed analytically. The values of $I_0$, $Q_0$, $U_0$ and $V_0$ are obtained by solving the
following $4 \times 4$ linear system:
\begin{equation}
\mathbf{A} \mathcal{I} = \mathbf{b},
\end{equation}
with 
\begin{equation}
\mathbf{A}=
\left[
\begin{array}{cccc}
\mathbf{N}_1 \cdot \mathbf{N}_1 & 0 & 0 & 0\\
0 & \mathbf{N}_2 \cdot \mathbf{N}_2 &  \mathbf{N}_2 \cdot \mathbf{N}_3 &  \mathbf{N}_2 \cdot \mathbf{N}_4 \\
0 & \mathbf{N}_3 \cdot \mathbf{N}_2 &  \mathbf{N}_3 \cdot \mathbf{N}_3 &  \mathbf{N}_3 \cdot \mathbf{N}_4 \\
0 & \mathbf{N}_4 \cdot \mathbf{N}_2 &  \mathbf{N}_4 \cdot \mathbf{N}_3 &  \mathbf{N}_4 \cdot \mathbf{N}_4
\end{array}
\right],
\end{equation}
and
\begin{equation}
\mathbf{b}=
\frac{1}{2}
\left[
\begin{array}{c}
\mathbf{N}_1 \cdot \left( \mathbf{S}_1^\mathrm{obs} + \mathbf{S}_2^\mathrm{obs} \right) \\
\mathbf{N}_2 \cdot \left( \mathbf{S}_1^\mathrm{obs} - \mathbf{S}_2^\mathrm{obs} \right) \\
\mathbf{N}_3 \cdot \left( \mathbf{S}_1^\mathrm{obs} - \mathbf{S}_2^\mathrm{obs} \right) \\
\mathbf{N}_4 \cdot \left( \mathbf{S}_1^\mathrm{obs} - \mathbf{S}_2^\mathrm{obs} \right)
\end{array}
\right],
\end{equation}
where
\begin{align}
\mathbf{N}_1 &= \mathbf{M}_1 \left(\mathbf{1}+\mathbf{F}^{-1} \alphabold \right) \nonumber \\
\mathbf{N}_2 &= \mathbf{M}_2 \left(\mathbf{1}+\mathbf{F}^{-1} \alphabold \right) \nonumber \\
\mathbf{N}_3 &= \mathbf{M}_3 \left(\mathbf{1}+\mathbf{F}^{-1} \alphabold \right) \nonumber \\
\mathbf{N}_4 &= \mathbf{M}_4 \left(\mathbf{1}+\mathbf{F}^{-1} \alphabold \right)
\end{align}
are vectors of length $N_\mathrm{meas}$.

\subsubsection{Optimization with respect to $\betabold$}
Finally, fixing $\alphabold$ and $\mathcal{I}$ results in a slightly more involved but still linear problem that can be solved analytically by
solving the following linear system:
\begin{equation}
\mathbf{A} \betabold = \mathbf{b},
\end{equation}
with
\begin{equation}
\mathbf{A} = \left[
\begin{array}{cccc}
I_0^2 \mathbf{P}_1 \cdot \mathbf{P}_1 & 0 & 0 & 0\\
0 & Q_0^2 \mathbf{P}_2 \cdot \mathbf{P}_2 & Q_0 U_0 \mathbf{P}_2 \cdot \mathbf{P}_3 & Q_0 V_0 \mathbf{P}_2 \cdot \mathbf{P}_4 \\
0 & U_0 Q_0 \mathbf{P}_3 \cdot \mathbf{P}_2 & U_0^2 \mathbf{P}_3 \cdot \mathbf{P}_3 & U_0 V_0 \mathbf{P}_3 \cdot \mathbf{P}_4 \\
0 & V_0 Q_0 \mathbf{P}_4 \cdot \mathbf{P}_2 & V_0 U_0 \mathbf{P}_4 \cdot \mathbf{P}_3 & V_0^2 \mathbf{P}_4 \cdot \mathbf{P}_4 
\end{array}
\right],
\end{equation}
and
\begin{equation}
\mathbf{b} = 
\left[
\begin{array}{c}
\frac{I_0}{2} \mathbf{P}_1 \cdot \left( \mathbf{S}_1^\mathrm{obs} + \mathbf{S}_2^\mathrm{obs} \right) - I_0^2 \mathbf{Q}_1 \cdot \mathbf{P}_1 \\
\frac{Q_0}{2} \mathbf{P}_2 \cdot \left( \mathbf{S}_1^\mathrm{obs} - \mathbf{S}_2^\mathrm{obs} \right) - \left(Q_0^2 \mathbf{Q}_2 - Q_0 U_0 \mathbf{Q}_3 - Q_0 V_0 \mathbf{Q}_3 \right) \cdot \mathbf{P}_2 \\
\frac{U_0}{2} \mathbf{P}_3 \cdot \left( \mathbf{S}_1^\mathrm{obs} - \mathbf{S}_2^\mathrm{obs} \right) - \left(U_0 Q_0 \mathbf{Q}_2 - U_0^2 \mathbf{Q}_3 - U_0 V_0 \mathbf{Q}_3 \right) \cdot \mathbf{P}_3 \\
\frac{V_0}{2} \mathbf{P}_4 \cdot \left( \mathbf{S}_1^\mathrm{obs} - \mathbf{S}_2^\mathrm{obs} \right) - \left(V_0 Q_0 \mathbf{Q}_2 - V_0 U_0 \mathbf{Q}_3 - V_0^2 \mathbf{Q}_3 \right) \cdot \mathbf{P}_4
\end{array}
\right],
\end{equation}
where we have made the following simplifications for the sake of a less crowded notation:
\begin{align}
\mathbf{P}_1 &= \mathbf{M}_1 \mathbf{F}^{-1} \alphabold, \qquad \mathbf{Q}_1 = \mathbf{M}_1 \mathbf{1} \nonumber \\
\mathbf{P}_2 &= \mathbf{M}_2 \mathbf{F}^{-1} \alphabold, \qquad \mathbf{Q}_2 = \mathbf{M}_2 \mathbf{1} \nonumber \\
\mathbf{P}_3 &= \mathbf{M}_3 \mathbf{F}^{-1} \alphabold, \qquad \mathbf{Q}_3 = \mathbf{M}_3 \mathbf{1} \nonumber \\
\mathbf{P}_4 &= \mathbf{M}_4 \mathbf{F}^{-1} \alphabold, \qquad \mathbf{Q}_4 = \mathbf{M}_4 \mathbf{1}.
\end{align}

\begin{figure*}
\includegraphics[width=\textwidth]{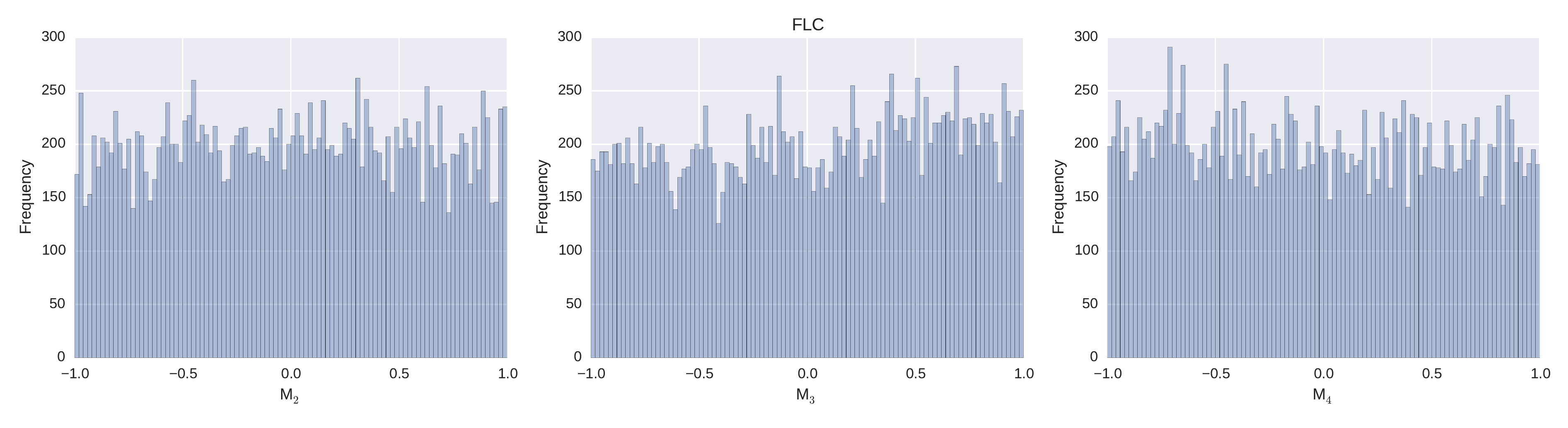}
\includegraphics[width=\textwidth]{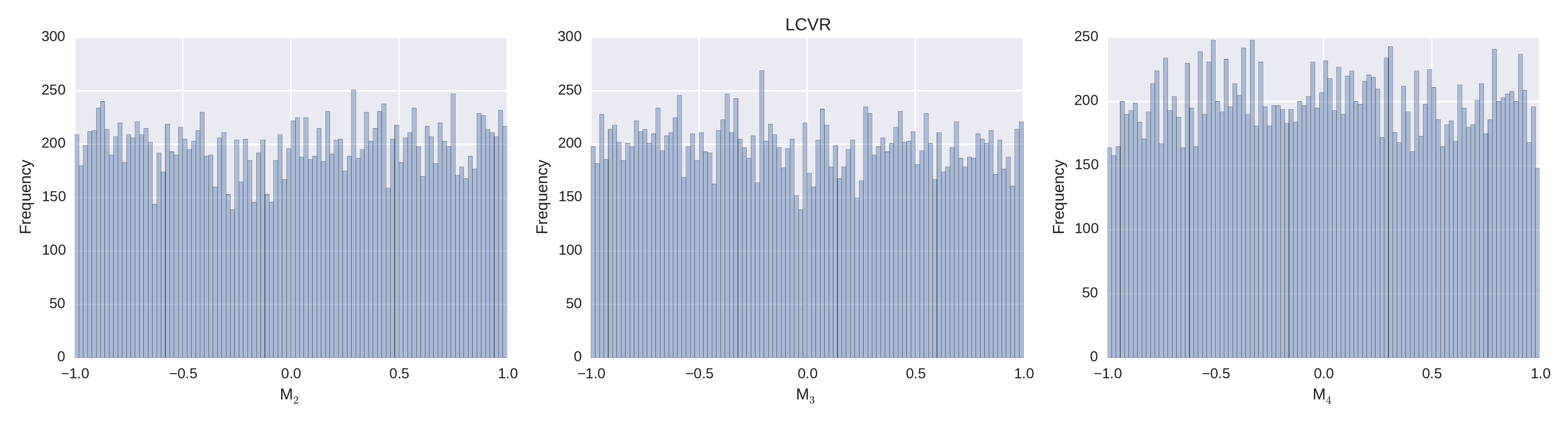}
\caption{Marginal distribution of the elements of the modulation matrix when the sampling is assumed to be
uniform. The upper panels show the results for two FLCs using Eq. (\ref{eq:modulationFLC}) and the lower
panels show the results for two LCVRs, using Eq. (\ref{eq:modulationLCVR}).}
\label{fig:modulationUniform}
\end{figure*}

\section{Examples}
We demonstrate the capabilities of the modulator we have described in the previous section with some synthetic examples.

\subsection{Random modulation matrices}
The first step is to define the modulation matrices we use in the design of the polarimeter.
We have verified that a suitable (if not the best) selection for the modulation matrices of Eqs. (\ref{eq:linearSystemMeasurements})
is to extract their matrix elements completely at random from the interval $[-1,1]$. The fundamental reason for this 
election is that such a random process is incoherent (cannot be efficiently developed) with the Fourier basis \citep{candes07,asensio_lopez_cs10} and this
helps extracting the Fourier coefficients of the signal by mixing all modes.

Assume a general polarimeter made of a train of two retarders and a beamsplitter. From the
options available in the market, we consider two different cases. First, we consider the case of FLC, that have a fixed retardance and can change the fast axis using different voltages. We note that commercial
FLC cannot change the fast axis at will, but only between two different values. If the retardance 
of the first retarder is set to $\pi/2$ so that it works as a $\lambda/4$ retarder plate and we set that of the second to $\pi$ so that
it behaves as a $\lambda/2$ retarder plate, the elements of the modulation matrix are (using standard Mueller algebra) given by:
\begin{align}
M_1(t) &= 1 \nonumber \\
M_2(t) &= \cos 2\alpha \cos \left[ 2 (\alpha-2\beta) \right] \nonumber\\
M_3(t) &= \sin 2\alpha \cos \left[ 2 (\alpha-2\beta) \right] \nonumber\\
M_4(t) &= \sin \left[ 2 (\alpha-2\beta) \right],
\label{eq:modulationFLC}
\end{align}
where $\alpha$ and $\beta$ are the angles defining the fast axis of the $\lambda/4$ and the $\lambda/2$, respectively.
The statistical properties of the elements of the modulation matrix are displayed in Fig. \ref{fig:modulation}
when $\alpha$ and $\beta$ are extracted randomly from a uniform distribution in the interval $[0,\pi/2]$. Note that
the modulation matrices do not allow to fill the full $[-1,1]^3$ cube because they fulfill the 
equality \citep[e.g.,][]{delToro07}:
\begin{equation}
\sqrt{M_2(t)^2+M_3(t)^2+M_4(t)^2} = 1,
\end{equation}
which describes the surface of a sphere. 

The elements of the modulation matrix are clearly correlated when the angles of the fast axis are
chosen randomly, we find that $M_2 \approx 0$ and $M_3 \approx 0$ are more probable, while $|M_4| \approx 1$ is
more probable. We have verified that the presence of this correlation negatively affects
the reconstruction algorithm described in Sec. 2C. A purely uniform random distribution in the elements of
the modulation matrix gives much better results. To this end, one needs to force 
the sampling to be uniform in the solid angle subtended by the three-dimensional hypersurface defined by Eqs. (\ref{eq:modulationFLC}).
Given that $\mathbf{r}(\alpha,\beta)=(M_2,M_3,M_4)$ is a unit vector, the solid angle is given by 
$\mathrm{d}\Omega = (\mathbf{n} \cdot \mathbf{r}) \mathrm{d} \alpha \mathrm{d} \beta$, where 
$\mathbf{n} = (\partial M_2/\partial \alpha,\partial M_3/\partial \alpha,\partial M_4/\partial \alpha) \times (\partial M_2/\partial \beta,\partial M_3/\partial \beta,\partial M_4/\partial \beta)$
is the normal vector. After some algebra, we find $\mathrm{d}\Omega = \cos[2(\alpha-2\beta)] \mathrm{d} \alpha \mathrm{d} \beta$. 
We use the \texttt{emcee} Python package \citep{emcee12}
to sample $\alpha$ and $\beta$ from this distribution. The upper panel of Fig. \ref{fig:modulationUniform} shows the probability distribution
for the elements of the modulation matrix, demonstrating that they are uniformly distributed.

\begin{figure*}[!t]
\includegraphics[width=\textwidth]{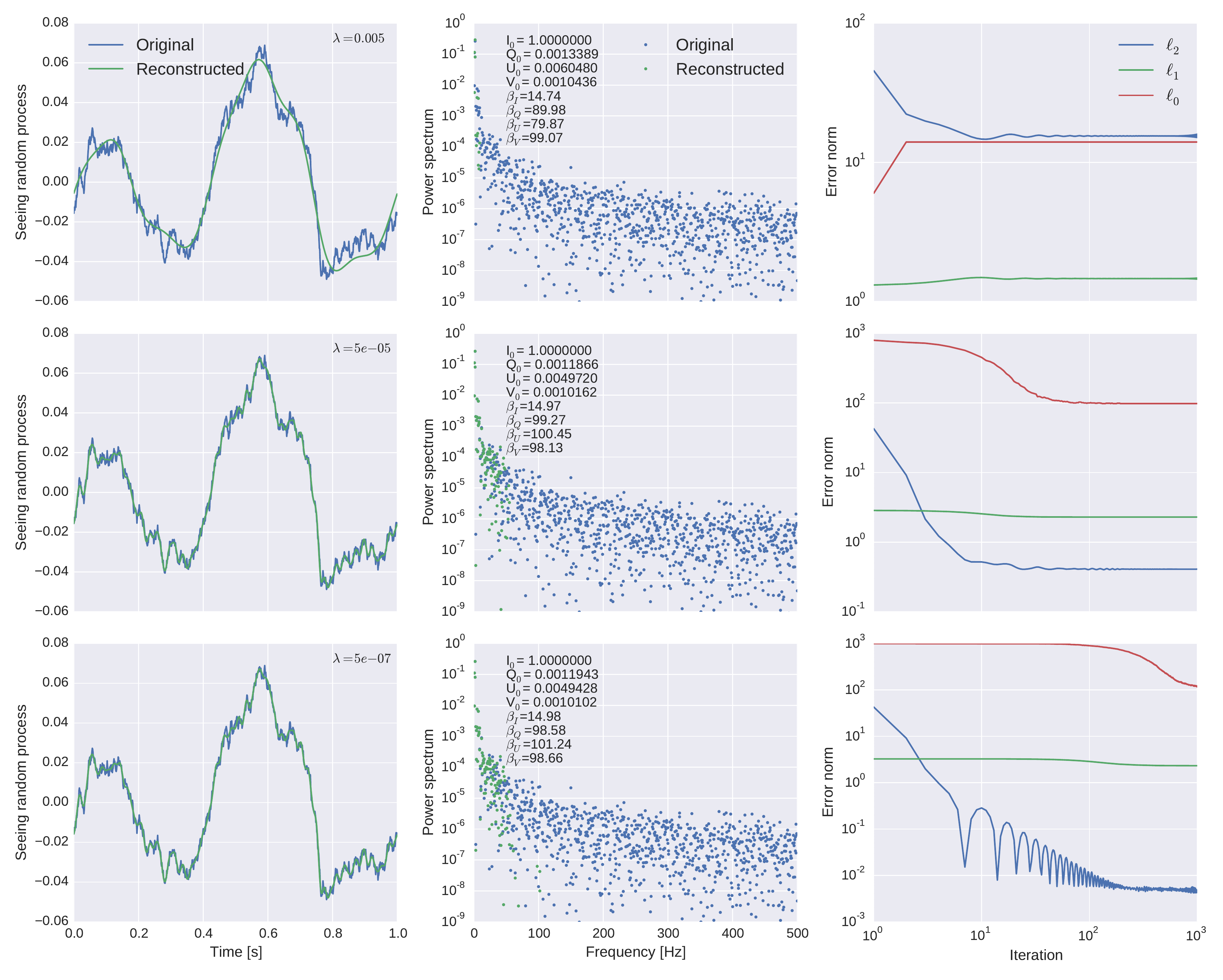}
\caption{Inferred high-frequency variation of the seeing for a total observing time of 1 s and an integration time of 10 ms for different values 
of the regularization parameter $\lambda$. The variation of the seeing is recovered down to 1 ms. The left panel shows the original seeing random process $N(t)$
in blue and the reconstructed process in green. The middle panel displays the ensuing power spectrum with the same color code. These middle panels also display
the inferred values of $\mathbf{I}$ and $\betabold$.
The right panel shows the evolution with iteration of the $\ell_0$ and $\ell_1$ norms of $\alphabold$, together with the $\ell_2$ norm 
of the residual, i.e., $\Vert g(\alphabold) \Vert_2$.}
\label{fig:noNoise}
\end{figure*}

The second case we consider is that of LCVR, optical elements that have a fixed fast axis and
a variable retardance, that can be modified in the range $[0,2\pi]$. If a polarimeter is built with two such retarders at 0$^\circ$ (with
a retardance $\delta_1$) and 45$^\circ$ (with a redardance $\delta_2$) plus a beamsplitter with one of the axis aligned
with the first retarder, the elements of the modulation matrix are given by:
\begin{align}
M_1(t) &= 1 \nonumber \\
M_2(t) &= \cos \delta_2 \nonumber\\
M_3(t) &= \sin \delta_1 \sin \delta_2 \nonumber\\
M_4(t) &= -\cos \delta_1 \sin \delta_2.
\label{eq:modulationLCVR}
\end{align}
Following the same approach as before, the lower panel of Fig. \ref{fig:modulationUniform} shows the distribution of elements of 
the modulation matrix, where now  $\mathrm{d}\Omega = \sin \delta_2 \mathrm{d} \delta_1 \mathrm{d} \delta_2$.
The results presented in the following sections are equivalent in the two cases given that uniformly distributed 
elements of the modulation matrix are used.

\subsection{Noiseless case}
In the following, we consider a total time of 1 s, a step $\Delta t=1$ ms and expositions with $\Delta T=10$ ms. Therefore,
we have $M=10$ and $N_\mathrm{meas}=100$. Therefore, we are considering solving the linear system of Eqs. (\ref{eq:linearSystemMeasurements})
that is underdetermined by a factor 10, which can be considered as a quite challenging aim. The 
Stokes parameters used in the simulations are $I_0=1$, $Q_0=1.2 \times 10^{-3}$, $U_0=5 \times 10^{-3}$ and $V_0=10^{-3}$, while
$\betabold=(10,150,150,150)$, in agreement with previous estimations \cite{judge04}.

\begin{figure*}
\includegraphics[width=\textwidth]{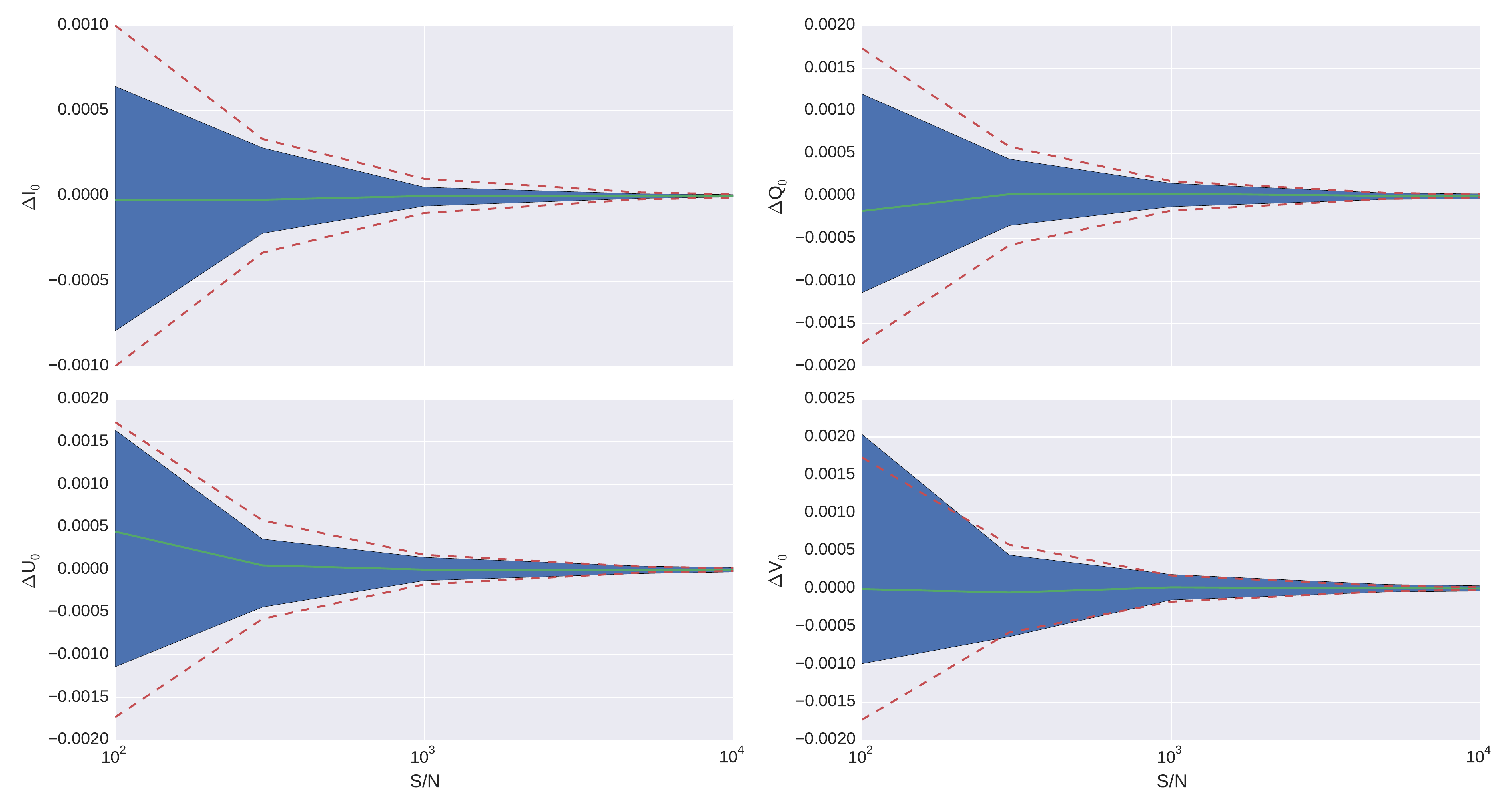}
\caption{Difference between the inferred Stokes parameters and the real one (e.g., $\Delta I_0=I_0^\mathrm{inferred}-I_0$). The shaded blue region marks the 1$\sigma$ region, with
the green line marking the median. The green dashed lines mark the inverse of the $S/N$ for individual measurements. The red dashed
lines display the inverse of the $S/N$ using all photons during the observing time of 1 s.}
\label{fig:noisySN}
\end{figure*}

We first consider the signal-to-noise ratio ($S/N$) to be extremely large, so that our measurements have technically no noise. Figure \ref{fig:noNoise}
considers the results for different values of the regularization parameter $\lambda$ (in each row). This parameter controls the sparsity of the solution. The 
largest its value, the sparsest the solution. When the value is too large, the inferred seeing random process $N(t)$ will be too smooth and the
temporal correlation will be lost. Otherwise, if the value is too small, the inferred random process will be that obtained by solving the linear 
system of Eqs. (\ref{eq:linearSystemMeasurements}) using a singular value decomposition. This least-squares solution tends to remove the temporal
correlation from the solution and will converge to pure Gaussian noise. Therefore, a compromise has to be found to obtain a good representation of 
the solution.

The left panels display the inferred time variation of the seeing random process $N(t)$ in green, together with the original
one in blue. The middle panels show the original power spectrum in blue dots while the reconstructed one is shown in green dots. This 
panel also shows the inferred value of $Q_0$, $U_0$ and $V_0$.
Finally, the right panels display the evolution during the iterative process of the $\ell_0$ (red) and $\ell_1$ (green) norms of $\alphabold$, i.e., 
$\Vert \alphabold\Vert_0$ and $\Vert \alphabold\Vert_1$, respectively, together with the $\ell_2$ norm of the residual, i.e.,
$\Vert \mathbf{S}_1 - \overline{\mathbf{S}}_1 \Vert_2^2 + \Vert \mathbf{S}_2 - \overline{\mathbf{S}}_2 \Vert_2^2$.
As described above, when the parameters $\lambda$ is too small, the inferred value of the seeing random process is very
smooth and the majority of the temporal correlations are lost. As a rule-of-thumb, when only $\sim$10\% of the Fourier coefficients are active,
we find a very good representation of the seeing random process and also a very good estimation of the value of the Stokes parameters.

\subsection{Noisy case}
Now that we have demonstrated that the method is able to recover the seeing random process and the value of the
Stokes parameters in an ideal situation without noise, we consider the more realistic 
case in which noise is injected during the observations. We consider the inferred value of $\mathcal{I}$ under the
same circumstances than the previous section but now adding a noise in each measurement resulting in a different $S/N$.
Figure \ref{fig:noisySN} displays the recovered values of $I_0$, $Q_0$, $U_0$ and $V_0$ for each value of the $S/N$. In order
to obtain information about the statistical properties of the recovered values, we have carried out a Montecarlo
analysis repeating 50 times each recovery of $\mathcal{I}$. Figure \ref{fig:noisySN} shows the difference between the recovered value of the Stokes parameters
and the real value for every $S/N$ in solid green lines, with the shaded blue region marking the values inside the percentiles 16 and 84 (that
delineates the $\pm 1\sigma$ region). Additionally, the dashed green lines indicate the expected values of the 
standard deviation for each $S/N$ for each individual measurement, while the dashed red lines display the expected
uncertainty in the ideal situation in which one could add up all $N_\mathrm{meas}$ to detect the signal.

\begin{figure}[!ht]
\includegraphics[width=\columnwidth]{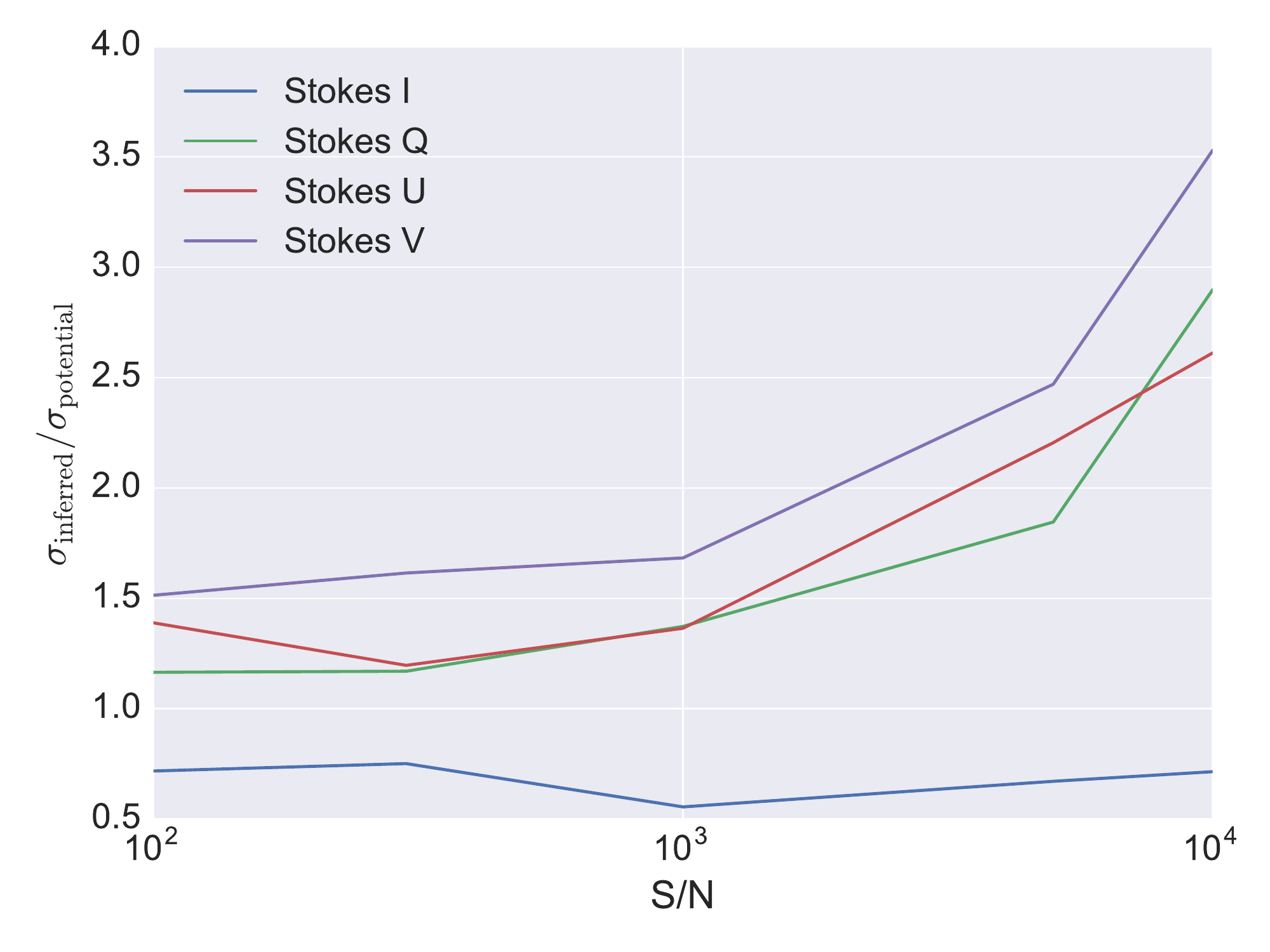}
\caption{Uncertainty in the value of the inferred Stokes parameters as compared with the one that would be
potentially obtained using the total integration time of 1 s if the seeing would be known with certainty.}
\label{fig:ratioSN}
\end{figure}

It is obvious that our reconstruction has a much better $S/N$ than that of individual measurements because we are
making use of many observations. However, we note that in Stokes $Q$, $U$ and $V$ we do not reach the limit of the $S/N$ that 
one would obtain in the ideal case that the seeing is known with certainty. In this case, one would just
add photons during the total integration time while compensating for the seeing. The fundamental reason for this is that 
some part of the acquired information goes into extracting the time variation of the seeing, which negatively
affects the inferred value of $\mathcal{I}$. This is shown in Fig. \ref{fig:ratioSN}, where we display the ratio between the uncertainty in the
Stokes parameters that we get with our reconstruction and the potential limit given by $(S/N) N_\mathrm{meas}^{1/2}$ (red dashed curves in Fig. \ref{fig:noisySN}) when using the
full integration time. The uncertainty associated with Stokes $I$ is equivalent to using the full integration time. However, this is
not the case for Stokes $Q$, $U$ and $V$. We get a factor $\sim 1.5$ increase in the uncertainty as compared with the potential
limit for $S/N \lesssim 10^3$, while it increases to $\sim 3$ for $S/N \sim 10^4$. This degradation is probably produced by the presence
of multiplicative photon noise.

\section{Conclusions}
We have theoretically demonstrated the feasibility of a polarimeter that modulates at the frequency of the seeing
but measures much slower. The recovery of the Stokes parameters and the time evolution of the seeing
is made possible by invoking the theory of compressed sensing using the fact that the Fourier power
spectrum of the seeing is weakly sparse. We have shown, using numerical simulations, the robustness of
the recovery under the presence of noise. We have described how polarimeters based on FLCs or LCVRs
can be tuned to deal with a highly efficient recovery of the Stokes parameters. The current 
technology in modulators start to be close to the requirements imposed by our approach. Extrapolating
the technological improvements in the last decades, we expect that the next generation of modulators will 
be compliant with the requirements.

The approach presented
here can work in the full-Stokes mode or it can also work for recovering only partial information 
with simplified polarimeters (for instance, in exoplanet work, where only IQU polarimetry is required).
Our approach to polarimetry is of interest until fast cameras integrating at the kHz can be used in
conjunction with fast modulators. 
Codes to reproduce the figures of this paper are available at \texttt{https://github.com/aasensio/randomModulator}.


\section*{Funding Information}

Financial support by the Spanish Ministry of Economy and Competitiveness 
through projects AYA2010--18029 (Solar Magnetism and Astrophysical Spectropolarimetry) and Consolider-Ingenio 2010 CSD2009-00038 
are gratefully acknowledged. 
AAR acknowledges financial support through the Ram\'on y Cajal fellowships. 

\section*{Acknowledgments}

I thank F. Snik and M. Collados for useful suggestions. This research has made use of NASA's
Astrophysics Data System Bibliographic Services.


\end{document}